\documentclass[journal]{IEEEtran}
\usepackage{xspace}
\usepackage{soul}
\usepackage{mathtools}
\usepackage{makecell}
\usepackage{cite}
\usepackage{nicefrac}
\usepackage{float}
\usepackage{wrapfig}
\usepackage{amssymb}
\usepackage{amsmath}
\usepackage[dvipsnames]{xcolor}
\usepackage{graphicx}

\usepackage{subcaption}
\usepackage[]{algorithm2e}
\usepackage{tikz}
\usetikzlibrary{spy,arrows}
\usepackage{pgfplots}
\pgfplotsset{compat = 1.10}

\newcommand{\ra}{\rightarrow}

\newcommand{\Ra}{\Rightarrow}

\newtheorem{dfn}{Definition}
\newtheorem{asm}{Assumption}
\newtheorem{prp}{Proposition}
\newtheorem{lem}{Lemma}
\newtheorem{thm}{Theorem}

\newtheorem{rem}{Remark}

\newcommand{\ie}{\unskip, i.\,e.,\xspace}
\newcommand{\eg}{\unskip, e.\,g.,\xspace}
\newcommand{\pd}{p.\,d.\xspace}

\newcommand{\wrt}{w.\,r.\,t.\xspace}

\newcommand{\N}{\ensuremath{\mathbb{N}}}

\newcommand{\R}{\ensuremath{\mathbb{R}}}

\newcommand{\T}{\ensuremath{\mathcal{T}}}
\newcommand{\X}{\ensuremath{\mathbb{X}}}

\newcommand{\F}{\ensuremath{\mathcal{F}}}
\newcommand{\U}{\ensuremath{\mathbb{U}}}

\newcommand{\eps}{\ensuremath{\varepsilon}}
\newcommand*\diff{\mathop{}\!\mathrm{d}}

\newcommand{\ball}{\ensuremath{\mathcal B}}
\newcommand{\K}{\ensuremath{\mathcal{K}}\xspace}		
\DeclareMathOperator*{\argmin}{arg\,min}

\DeclareMathOperator*{\arginf}{arg\,inf}

\newcommand{\blue}[1]{\textcolor{black}{#1}}

\newcommand{\spc}{\ensuremath{\,\,}}	

\makeatletter
\newcommand{\subalign}[1]{%
	\vcenter{%
		\Let@ \restore@math@cr \default@tag
		\baselineskip\fontdimen10 \scriptfont\tw@
		\advance\baselineskip\fontdimen12 \scriptfont\tw@
		\lineskip\thr@@\fontdimen8 \scriptfont\thr@@
		\lineskiplimit\lineskip
		\ialign{\hfil$\m@th\scriptstyle##$&$\m@th\scriptstyle{}##$\crcr
			#1\crcr
		}%
	}
}

\newcommand*{\Scale}[2][4]{\scalebox{#1}{$#2$}}%
\usepackage{ marvosym }
\usepackage[utf8]{inputenc}
\newcommand{\W}{\ensuremath{\mathbb{W}}}
\newcommand{\J}{\ensuremath{\mathbb{J}}}

\newcommand{\V}{\ensuremath{\mathcal{V}}}
\ifCLASSINFOpdf
\else
\fi
\begin{document}
	%
	\title{\blue{A stabilizing reinforcement learning approach for sampled systems with partially unknown models}}
	
	
	\author{Lukas~Beckenbach~and~Pavel~Osinenko~and~Stefan~Streif
		\thanks{Lukas~Beckenbach
			and~Stefan~Streif (Corr. Author) are with the Laboratory for Automatic Control and System Dynamics, Technische Universität Chemnitz, 09107 Chemnitz, Germany. Pavel Osinenko is with the Digital Engineering Center, Skolkovo Institute of Science and Technology, Moscow, Russia 121205.}
		\thanks{Email: \{lukas.beckenbach,stefan.streif\}@etit.tu-chemnitz.de, p.osinenko@skoltech.ru}%
	}
	
	\markboth{}%
	{Beckenbach, Osinenko \& Streif: A stabilizing reinforcement learning approach for sampled systems with partially unknown models}
	%
	
	
	
	\maketitle
	
	\begin{abstract}
		Reinforcement learning is commonly associated with training of reward-maximizing (or cost-minimizing) agents, in other words, controllers. 
		\blue{It can be applied in model-free or model-based fashion, using a priori or online collected system data to train involved parametric architectures. 
			In general, online reinforcement learning does not guarantee closed loop stability unless special measures are taken, for instance, through learning constraints or tailored training rules.}
		\blue{
			Particularly promising are hybrids of reinforcement learning with ``classical'' control approaches}.
		In this work, we suggest a method to guarantee practical stability of the system-controller closed loop in a purely online learning setting \ie without offline training.
		Moreover, we assume only partial knowledge of the system model.
		To achieve the claimed results, we employ techniques of classical adaptive control.
		The implementation of the overall control scheme is provided explicitly in a digital, sampled setting.
		That is, the controller receives the state of the system and computes the control action at discrete, specifically, equidistant moments in time.
		The method is tested in adaptive traction control \blue{and cruise control} where it proved to significantly reduce the cost.
	\end{abstract}
	
	
	
	
	%
	\IEEEpeerreviewmaketitle
	
	
	\section{Introduction}
	\label{sec:Introduction}
	
	
	Reinforcement learning is an optimal control method that employs learning techniques imitating adaptation of living beings in environments \cite{Barto2004,Sutton2018}.
	It entails a data-driven character through exploitation of the state space to infer an optimal policy, usually over an infinite horizon of future rewards (or stage costs, depending on the application), \blue{possibly with experience replay \cite{Adam2012}}.  
	Oftentimes, reinforcement learning implies extensive offline training of controllers to learn the optimal policy \blue{though there exist online learning approaches that omit offline state exploration}. 
	However, guaranteeing stability of the closed loop is a challenge for online reinforcement learning \blue{and requires careful selection of the training procedure}.
	Attempts have been made towards this problem \blue{using so-called adaptive critics \cite{Konda2003,Barto1991,Barto1995,Bertsekas2017}, also referred to as value-based methods,} using specially tailored learning rules.
	
	
	Oftentimes the existing approaches, which are reviewed in detail below, rely on prior assumptions about the optimal infinite-horizon cost (the optimal accumulated stage cost) and/or the optimal policy (see \eg \cite{Xiao2015,Mu2017}) \blue{or assume additional information about the system such as the existence of a control Lyapunov function \cite{Kim2020} or a fallback controller \cite{Mahmud2021}}.
	\blue{The current work takes the latter path and suggests to analyze closed loop stability of a model-based reinforcement learning design by means of a particular adaptive control framework, namely the use of an adaptive control Lyapunov function, and to ensure stability without resorting to assumptions on the infinite-horizon cost as well as to omit extensive offline data collection and controller training. 
	In comparison to \eg \cite{Kim2020}, the herein presented approach is an online learning design in which the controller is not bound to Sontag's formula \cite{Sontag1989}, requiring the infinite-horizon cost approximant to be a control Lyapunov function, and a specific condition on the approximant learning rate is proposed. }
	
	
	\subsubsection*{Related work}
	
	\paragraph{Safe Reinforcement Learning}
	In safe reinforcement learning, performance and stability includes the notion of risks \blue{which seeks to overcome risky \eg arbitrary, state space exploration and render it sensitive to risks such as safety constraint violation} (see \eg \cite{Garcia2015} for a survey). 
	Such safety aspects were also addressed \eg in \cite{Berkenkamp2017}, where a region of attraction to a given safe policy is characterized and \blue{a policy optimization procedure including safe state space exploration is proposed}. 
	A related investigation was made in \cite{Richards2018}, where a neural network Lyapunov function candidate is used to construct sets for safe control learning. 
	In \cite{Koller2018}, safety and stability is provided through multi-step ahead predictions in combination with safety constraints, relying on a given backup controller. 
	The importance of a (local) backup controller, which is frequently used in \eg robust and learning-based model predictive control, is highlighted in several studies \cite{Koller2018,Nubert2020,Koehler2021a}.  
	\blue{It is noteworthy that the aforementioned analyses use a system model \ie are model based, to assert closed loop properties.} 
	Early works of (robustly) stabilizing reinforcement learning include \eg \cite{Kretchmar2001} in which integral quadratic constraints (IQCs) were introduced to inspect closed loop stability properties. 
	\blue{ICQs allow learning related errors in the closed loop to be treated as bounded (input) uncertainty} and have appeared in several other stability analyses including \eg the recent work \cite{Jin2020}. 
	
	\paragraph{Model-free Reinforcement Learning}
	\blue{Complementary to model-based analyses, so-called ``model-free'' control problems \cite{Vamvoudakis2017,Sun2020,Luo2017,Xiao2015,Fazel2018} or such with partially unknown dynamics \cite{Dierks2012,Mahmud2021} have been gaining interest. 
		The unknown part may \eg be approximated by a parametric structure after which a model-based analysis can be conducted \cite{Xiao2015,Liu2012,Bhasin2013}.} 
		Remaining model uncertainties which may be interpreted as external disturbances were investigated in \eg \cite{Zhang2017,Zhang2018} or handled by the aforementioned ICQs in \cite{Jin2020}. 
	\blue{Integral reinforcement learning may even omit requiring (and thus modeling) the systems drift dynamics (see \eg \cite{Modares2014})}. 
	Though the control design may be model free, certain assumptions on the dynamics are required to ensure closed loop properties such as \eg second-order differentiability \cite{Bhasin2013} or linearity \cite{Vamvoudakis2017} on the uncertain dynamics. 
	As was noted in \eg \cite{Berkenkamp2017}, such assumptions are critical in particular for safety aspects.

	\paragraph{Convergence Results}
	For either model-based or model-free online reinforcement learning, exploration of the state space is desirable when addressing convergence of controller and/or cost approximants to their respective best fit under a selected approximating structure. 
	Ultimate uniform boundedness (UUB) denotes one specific notion of stability which can be found frequently for controller and cost approximant parameters (as well as possibly the closed loop state) \cite{Vamvoudakis2014,Vamvoudakis2016,Modares2013,Xiao2015,Dierks2009}. 
	\blue{A repeatedly observable property (in preceding listed references) is the dependence of convergence bounds on the choice of approximant and related approximation errors. 
		Though adjustable through offline training experience, it may be unsuitable for online learning approaches on problems.} 
	
	
	In conclusion, closed loop stability is a result of specific construction of the controller's learning rule as well as that of the cost approximation; most commonly gradient or optimization based. 
	\blue{While offline trained approximants, remaining fixed during the de facto control process (refer to \eg \cite{Sokolov2015} and references therein), may offer more possibilities in state space exploration, (purely) online learning controllers base experience on the current trajectory. 
		Subsequently, risky \eg stability harming, control actions must be carefully addressed and still requires significant research efforts.}  
	
	\paragraph{Adaptive Control} The field of adaptive control \cite{Krstic1995,Sastry2011} offers a variety of control strategies in which (parametrically) unknown dynamics can be integrated into the reinforcement learning setting and vice versa by augmenting the controller with an identification apparatus. 
	Asymptotic stability can be achieved under various techniques, one of which follows control and system identification rules alongside an adaptive control Lyapunov function \cite{Krstic1995a}. 
	Although it can be argued that finding such an adaptive control Lyapunov function is potentially hard, this approach does not require convergent parameters (to the true parameters) in the model or of the cost approximation to stabilize the system. 
	Some effort has already been made to improve adaptive control approaches by incorporating optimal control components in the form of (near) optimal value functions \cite{Lopez2021}.
	However, no respective fusion of adaptive control Lyapunov function based control and reinforcement learning has been made in the literature to the knowledge of the authors, which motivates the current work. 
	
	
	\subsubsection*{Contribution}
	
	This work suggests a new model based reinforcement learning method with closed-loop stability guarantee for nonlinear systems with parametric uncertainty. 
	We start with the classical techniques of adaptive control that do not require convergence of the parameter estimate, design special learning constraints and show that the resulting controller practically stabilize the system in the sample-and-hold sense.
	That is, the reinforcement learning controller updates neural network weights and computes actions at discrete equidistant moments in time.
	\blue{The main result may be summarized as follows:}
	\begin{itemize}
		\item[] \blue{The designed reinforcement learning controller practically stabilizes the given system with parametric uncertainty under the designed learning conditions, provided adaptive control Lyapunov function with an associated stabilizing controller.}
	\end{itemize}

	Practical stability commonly arises from consideration of system properties in sampled settings \cite{Clarke1997-stabilization}. 
	\blue{The controller and cost approximant, which are sought to satisfy these stability conditions, are trained online via reinforcement learning rules. 
	An adaptive control Lyapunov function is assumed to be known for the reinforcement learning control design}.  
	\blue{The stability conditions that are posed as learning conditions are checked at each sampling instance, though they may be embedded into the related optimization routine as constraints. 
	If a violation of these conditions is detected, the nominal stabilizing controller associated with the adaptive control Lyapunov function is invoked.}
	By this approach, practical stability of the closed loop is assured.
	
	
	
	
	\blue{In the following Section~\ref{sec:preliminaries}, preliminary tools are introduced including the adaptive control framework used to embed reinforcement learning mechanisms reviewed thereafter. 
		Furthermore, the sample-and-hold approach is reviewed leading along with practical stability. 
		Then, based on the foregoing insight, a learning based adaptive controller algorithm is proposed.
		Section~\ref{sec:main-results} covers the closed loop analysis and concludes closed loop practical stability. 
		Two case studies in Section~\ref{sec:case-study} demonstrate the efficacy of the approach on processes to which an adaptive control Lyapunov function is available.}
	
	
	
	\subsubsection*{Notation} A closed ball of radius $r>0$ centered at $z \in \R^n$ is denoted by $\ball_r(z)$, while $\ball_r := \ball_r(0)$ for brevity. 
	The inner product is denoted as $\langle \bullet,\bullet \rangle$. 
	For any $z \in \R^n$ and positive semi-definite $H \in \R^{n \times n}$, $\|z\|$ denotes the two-norm 
	while $\|H\|$ \blue{denotes a matrix norm} and $\lambda_{\text{min}}(H)$ as well as $\lambda_{\text{max}}(H)$ the smallest and largest eigenvalue, respectively. 
	\blue{The partial derivative of $J(z) \in \R$ with respect to $z \in \R^n$ is denoted by $\nabla_z J(z) = \partial J(z) / \partial z$, being a column vector}.



	\section{Preliminaries}
	\label{sec:preliminaries}
	
	This section introduces the problem setup as well as certain related and utilized tools to solve the former. 
	
	\subsection{Problem Setup}
	
	
	This work is concerned with nonlinear dynamical systems 
	\begin{equation} \label{eq:sys}
		\begin{array}{rl}
			\dot{x} & = \; f(x) + F(x) \theta + g(x) u, \, x(0)  = x_0, \, t \geq 0,
		\end{array}
	\end{equation}
	with state $x \in \R^n$, input $u \in \R^{m}$ and parameter $\theta \in \R^p$, where $f:\R^n \ra \R^n,g:\R^n \ra \R^{n \times m},F:\R^n \ra \R^{n \times p}$ are continuously differentiable and $f(0) = 0,F(0) = 0$. 
	For brevity, denote $\mathcal{F}(x,u,\theta) \coloneqq f(x) + F(x) \theta + g(x) u$.  
	
	The control task motivating the later proposed approach is to find a control $u(\cdot)$ that minimizes 
	\begin{align} \label{eq:IH-cost}
		J(x_0,u(\cdot)) \coloneqq \int \limits_{0}^{\infty} r(x(t),u(t)) \, \diff t
	\end{align}
	along the solution $x(\cdot)$ to \eqref{eq:sys} under $u(\cdot)$ for any $x_0$, where $r:\R^n \times \R^m \ra \R_{\geq 0}$ is a continuous positive-definite (\pd) \textit{running cost} \ie satisfying $r(0,0)=0$ and $r(x,u)>0$ otherwise. 
	Denote the optimal cost $J^\ast(x) \coloneqq  J(x,u^\ast(\cdot))$ with the associated minimizing control policy $u^\ast(\cdot)$. 
	
	Under the Bellman's principle of optimality, $J^\ast$ satisfies the Hamilton-Jacobi-Bellman (HJB) equation
	\begin{align} \label{eq:HJB}
		0 = \min_{u \in \R^m} \; r(x,u) + \nabla_x^\top J^\ast(x) \mathcal{F}(x,u,\theta)
	\end{align}
	for all $x \in \R^n$ (assuming the minimum exists). 
	Due to the curse of dimensions, finding $J^\ast$ \blue{analytically} is challenging \cite{Liberzon2011}, though certain techniques exist to compute the latter approximately on $\R^n$. 
	
	Herein, the parameter $\theta$ is assumed unknown which poses a challenge and will be addressed via adaptive control.

%
	\subsection{Adaptive Control Lyapunov Function}
	\label{subsec:ACLF}
	
	\blue{The adaptive control literature offers a great variety of tackling partially or completely unknown system dynamics; one particular approach involves an adaptive control Lyapunov function as an extension of the classical control Lyapunov functions \cite{Sontag1989}. 
	}  
	
	
	\begin{dfn}[Adaptive CLF \cite{Krstic1995a}] \label{def:ACLF}
		A smooth function $V:\R^n \times \R^p \ra \R_{\geq 0}$, \pd and proper in $x$ and $\theta$ is called an adaptive control Lyapunov function for \eqref{eq:sys} if there exists a \pd matrix $\Gamma \in \R^{p \times p}$ such that for each $\theta \in \R^p$, $V(x,\theta)$ is a CLF for the modified system
		\begin{align} \label{eq:sys-mod}
			\begin{split}
				\dot{x} = \underbrace{f(x) + F(x) \left( \theta + \Gamma \, \nabla_{\theta}V(x,\theta)  \right) + g(x)u}_ {=: G(x,u,\theta)}.
			\end{split}
		\end{align}
		That is, for any bounded subset $\X \subset \R^n$ containing the origin there exists a compact set $\U_{\X} \subset \R^m$ and a continuous function $\nu:\R^n \times \R^p \ra \R_{\geq 0}$, \pd in $x$ for all $\theta$, such that 
		\begin{align}
			\begin{split}
				&\forall \theta \in \R^p, \, \forall x \in \X, \, \exists u \in \U_{\X}, \\
				&\left\langle \nabla_x V(x,\theta), G(x,u,\theta) \right\rangle \leq -\nu(x,\theta).
			\end{split}
		\end{align}
	\end{dfn}
	
	Given $V$, a policy $u=\mu(x,\theta):\R^n \times \R^p \ra \R^m$ can be constructed according to \eg Sontag's formula \cite{Sontag1989}. 
	
	Furthermore, for $\Theta \in \R^p$, there exists $\alpha_{1,2}^{\theta} \in \mathcal{K}_{\infty}$ such that $\alpha_1^{\theta}(\|x\|) \leq V(x,\theta) \leq \alpha_2^{\theta}(\|x\|)$ for all $\theta \in \Theta$. 
	
	
	
	
	Along the control process only an estimate $\hat{\theta} = \hat{\theta}(t)$ of the unknown $\theta$ is available. 
	Within the adaptive control Lyapunov function framework it is suggested to use
	\begin{align} \label{eq:param-estim-dynam}
		\begin{split}
			\dot{\hat{\theta}} = \Gamma\tau(x,\hat{\theta}), \; \; 
			\tau(x,\hat{\theta}) =  \left( \nabla_x^\top V(x,\hat{\theta}) F(x)  \right)^\top.
		\end{split}
	\end{align}
	\begin{lem}[\hspace*{-3pt}\cite{Krstic1995a}] \label{lem:LF-V_c}
		Let $V$ be an adaptive control Lyapunov function as per Def. \ref{def:ACLF} and let 
		\begin{align} \label{eq:V_c-LF}
			V_{c}(x,\tilde{\theta}) \coloneqq V(x,\hat{\theta}) + \frac{1}{2} \tilde{\theta}^\top \Gamma^{-1} \tilde{\theta},
		\end{align}
		with $\tilde{\theta}:= \theta- \hat{\theta}$, $\Gamma \in \R^{p \times p}$ \pd. 
		Then 
		\begin{align} \label{eq:decay-V_c}
			\dot{V}_{c}(x,\tilde{\theta}) \leq - \nu(x,\hat{\theta}) 
		\end{align}
		along the solution of \eqref{eq:sys} under $u=\mu(x,\hat{\theta})$ and \eqref{eq:param-estim-dynam}. 
	\end{lem}
	
	It follows that $x=0$ and $\hat{\theta} = \theta$ are globally stable and furthermore $x \ra 0$ \cite{Krstic1995a}. 
	Note that the decay rate $\nu$ of the adaptive control Lyapunov function $V$ is recovered in $V_c$.

	The next assumption is central to the current work.
	\begin{asm} \label{asm:ACLF-exist}
		System \eqref{eq:sys} admits a smooth adaptive control Lyapunov function $V:\R^n \ra \R_{\geq 0}$ as per Def. \ref{def:ACLF}, 
		for which there exist $\alpha_{1,2} \in \mathcal{K}_{\infty}$ and $\alpha_{\nu} \in \mathcal{K}$ such that $\alpha_1(\|x\|) \leq V(x) \leq \alpha_2(\|x\|)$ and $\alpha_{\nu}(\|x\|) \leq \nu(x,\theta)$ for all $(x,\theta) \in \R^n \times \R^p$ \ie
		\begin{align} \label{eq:ACLF-decay}
			\left\langle \nabla_x V(x), G(x,u,\theta) \right\rangle \leq - \nu(x,\theta) \leq -\alpha_{\nu}(\|x\|).
		\end{align}
	\end{asm}
	
	\begin{rem} \label{rem:ACLF-theta-indepentent-on-compact}
		For any compact $\Theta$, one can always choose $\alpha_{1,2} \in \mathcal{K}_{\infty}$ independent of $\theta$ such that for all $\theta \in \Theta$, $\alpha_1(\|x\|) \leq V(x,\theta) \leq \alpha_2(\|x\|)$ for any $x$, whence explicit dependence on $\theta$ in $V$ is omitted for the ease of exposition.
	\end{rem}
	
	Since $V$ is independent of $\theta$, $\tau(x,\theta) = \tau(x)$ in \eqref{eq:param-estim-dynam}. 
	
	\begin{rem}
		\blue{From a practical viewpoint, requiring knowledge of a (parameter independent) adaptive control Lyapunov function may be appear challenging to satisfy, but luckily a great variety of concrete examples exists. 
		These include \eg the adaptive cruise control \cite{Ames2014,Taylor2020}, car traction control \cite{Nakakuki2008} and biochemical processes \cite[Chap. 3.4.2]{Krstic1995}. 
		Also, systematic procedures exist to construct adaptive control Lyapunov functions for systems in the strict-feedback form.}
	\end{rem}
	
	
	
	\subsection{Sample-and-Hold Framework}
	\label{subsec:sample-and-hold}
	
	Consider the following sample-and-hold (SH) form of \eqref{eq:sys}
	
	\begin{align} \label{eq:sys-SH}
		\begin{split}
			&\dot{x} = \mathcal{F}(x,u_k,\theta), \quad x(0), \\
			&t \in [k \delta, (k+1) \delta), \quad u_k = \kappa(x(k \delta)), \quad k \in \N_0.
		\end{split}
	\end{align}
	At any $k \in \N_0$ and sampling period $\delta >0$, the state trajectory $x(t;u)$, $t \in [0,\delta )$, under any $u \in \R^m$ is defined as
	\begin{align} \label{eq:state-trajectory-explicit}
		x(t;u) = x_k + \int_{0}^{t} \mathcal{F}(x(k\delta + s),u,\theta) \; \text{d}s.
	\end{align}
	Denote $x_k := x(k\delta)$ and $x_{k+1}(u) \coloneqq x((k+1) \delta;u)$, for brevity and the corresponding trajectory to \eqref{eq:sys-SH} will be referred to as the SH-trajectory.
	
	\begin{dfn} \label{def:pract-stab}
		A control policy $\kappa(\cdot)$ is said to practically semi-globally stabilize \eqref{eq:sys} if, given $0 < r < R < \infty$, there exists a $\bar{\delta}>0$ such that any SH-trajectory of \eqref{eq:sys-SH} with sampling time $0 < \delta \leq \bar{\delta}$, starting in $\ball_R$, is bounded, enters $\ball_r$ after a time $T$, which depends uniformly on $R,r$, and remains there for all $t \geq T$. 
	\end{dfn}

	
	In the sampled setting, the next section introduces the model-based learning approach to be utilized herein.

	\subsection{A Model-Based Reinforcement Learning Method}
	\label{subsec:ADP}
	
	
	In the following, some elements of a particular reinforcement learning inspired method, namely approximate dynamic programming (ADP), are introduced that will be used in the presented control scheme (see Sec. \ref{subsec:actor-critic}). 
	\blue{Since $J^\ast$ is difficult to determine analytically and thus the associated optimal controller is hard to obtain}, a parametric model $\hat{J}= \hat{J}(x,w)$ called ``critic'' is employed to approximate $J^\ast$ \eg by the use of neural networks (see \eg \cite{Lewis2009,Wang2009,Powell2007} and references therein). 
	\blue{An approximate optimal controller can be drawn from \eg the HJB \eqref{eq:HJB} with $\hat{J}$ replacing $J^\ast$. }
	
	\begin{rem}
		\blue{A particular difficulty comes from the fact that $\theta$ is unknown and a substitute $\hat{\theta}$ must be used in the minimization of \eqref{eq:HJB}. 
		Adaptive control aims at tuning such a $\hat{\theta}$ so as to achieve the closed-loop stability.
		This perk of adaptive control is made use of in the currently suggested reinforcement learning method.
		} 
	\end{rem}
	
	
	Define the critic as 
	\begin{align} \label{eq:critic}
		\hat{J}(x,w): \R^n \times \R^l \ra \R_{\geq 0}; (x,w) \mapsto \langle w,\varphi(x) \rangle,
	\end{align} 
	with weights $w \in \R^l$, $l \in \N$ and $\varphi:\R^n \ra \R^l$ is the so-called regressor, consisting of $l$ \blue{twice} continuously differentiable basis functions, that satisfies the local Lipschitz continuity property: for any $0<r<\infty$ and $z \in \R^n$, $\exists L_{\varphi}(z,r)>0$ such that for all $x,y \in \ball_{r}(z)$,
	\begin{align} \label{eq:Lipschitz-activ-fcn-x}
		\| \varphi(x) - \varphi(y) \| \leq L_{\varphi}(z,r) \cdot \|x-y \|.  
	\end{align}
	
	The choice \eqref{eq:critic} 
	has been used frequently in the literature (see \eg \cite{Vamvoudakis2010,Wang2014,Jiang2019}). 
	Note that, as an approximant of the cost $J^\ast(x)$, the \emph{critic is constructed independently of $\theta$ or its estimate $\hat{\theta}$}.
	
	Policy iteration \cite{Vrabie2009a,Liu2014} denotes one particular learning procedure to improve upon the controller performance by iteratively solving HJB type equations for a controller $u$ and weights $w$. 
	For instance, given a controller $u(x)$, the weight $w^{i}$ of an iteration step $i\in \N_0$ is sought to solve 
	\begin{align} \label{eq:critic-opt-theoretic}
		\blue{w^\top\varphi(x) = \underbrace{r(x,u(x))  + w^\top\varphi(x + \delta\mathcal{F}(x,u(x),\hat{\theta})) }_{=:\mathbb{J}(x,u,w,\hat{\theta})} }
	\end{align} 
	for $w$, for all $x \in \R^n$ \blue{with, ideally, $\hat{\theta} = \theta$}. 
	In practice, this may be approached by the least-squares solution as per
	\begin{align} \label{eq:critic-opt-practical}
		\blue{w = \arginf_{v \in \R^l} \;\sum_{j=1}^M \Big( v^\top\varphi(x^j) - \mathbb{J}(x^j,u(x^j),v,\hat{\theta}) \Big)^2 }
	\end{align} 
	\blue{over $M$ samples $x^j \in \R^n$. 
		To ensure existence of $w$, $v \in \W \coloneqq \{w \in \R^l: \underline{w} \leq \|w\| \leq \overline{w} \}$, $\underline{w},\overline{w}>0$, may be employed as constraint. }
	\blue{Another common option is to adapt the critic weights via gradient rules roughly in the following form (refer to \eg \cite{Sokolov2015,Luo2017})
		\begin{align} \label{eq:GD-update-u-w}
			w_{k} = w_{k-1} - l_w \nabla_w (\hat{J} - \mathbb{J}), k \in \N_0,
		\end{align}
		with suitably chosen learning rates $l_{w}>0$.}
	 
		\blue{Subsequently, $\|w_{k} - w_{k-1}\| \leq \eps_w$ for some $\eps_w \geq 0$. }
	
	Based on \eqref{eq:critic-opt-theoretic}, 
	\blue{the approximate optimal control may be found by gradient methods analogously, adjusting weights of a parameterized $u_k = u_k(\varsigma_k)$ (cf. policy gradient methods \cite{Silver2014}) in the general form \cite{Sokolov2015}
	\begin{align} \label{eq:GD-input-params}
		\varsigma_k = \varsigma_{k-1} - l_{\varsigma}\nabla_{\varsigma} \mathbb{J},
	\end{align}
	with $l_{\varsigma}>0$ or} via optimization as per
	\begin{align} \label{eq:actor-critic-opt}
		\begin{split}
			\hspace*{-2em}\mathcal{P}(x_k): \quad  
				u_k = &\argmin_{u \in \U_{\X}} \quad  \blue{\J(x_k,u,w_{k},\hat{\theta}_k)}. 
			\end{split}
		\end{align}
	Here $\U_{\X}$ is the control set associated with the adaptive control Lyapunov function. 
	
	Subsequently, $\|u_k - u_{k-1}\| \leq \eps_u$ for some $\eps_u \geq 0$ associated to $\eps_w$.

	\begin{rem} \label{rem:online-RL-objective}
		Note that in online reinforcement learning, repeated runs are not available and thus finding $w$ at any time $t$ can either depend only on $x(t)$ or a history of $M$ state samples. 
		The use of data has impact on the performance of $\hat{u}$ which may be estimated empirically. 
		However, the focus herein lies in ensuring that $\hat{u}$ (practically) stabilizes the system despite errors in $\|w^\ast - w\|$ or $\|\theta - \hat{\theta}\|$. 
	\end{rem}

	
	
	Consider the user-defined $\varphi$ to satisfy the following:
	\begin{asm} \label{asm:struct-equiv}
		For any bounded $\X \subset \R^n$ containing the origin, i) there exist $q_1,q_2 \in \mathcal{K}_{\infty}$ such that for all $x \in \X$ and all $w \in \W$, $ q_1(\|x\|) \leq \hat{J}(x,w) \leq q_2(\|x\|)$, ii) there exist $\bar{\varphi}_1,\bar{\varphi}_2 >0$ such that 
		\begin{align*}
			\sup_{\subalign{x &\in \X \\ w &\in \W}} \; \| \nabla_x \hat{J}(x,w) \| \leq \bar{\varphi}_1, \; \; 
				\sup_{\subalign{x &\in \X \\ w &\in \W}} \; \| \nabla_x^2 \hat{J}(x,w) \| \leq \bar{\varphi}_2,
		\end{align*}
		iii) given an adaptive control Lyapunov function $V(x)$, there exists $w^{\#} \in \W$ such that
		\begin{align} \label{eq:ACLF-critic-equiv}
			V(x) \equiv \hat{J}(x, w^{\#}),
		\end{align}
		in other words, $\forall x \in \X \spc V(x) = \hat{J}(x, w)$,
		\blue{and, iv), there exists a subset $\W' \subset \W$ such that for all $w \in \W'$, $V(x) \leq \hat{J}(x,w)$ for all $x \in \X$. }
	\end{asm}
	
	
	\begin{rem}
		All requirements of Assumption \ref{asm:struct-equiv} can be met by suitable design of $\varphi$. 
		For instance, iii) implies that the activation function be constructed so as to allow structural matching with $V$, which is thus required in analytical form, through a particular $w$. 
		This assumption may be relaxed in principle to approximate structure matching without disrupting the subsequent analysis substantially. 
		\blue{By construction of $\varphi$, the set $\W$ may be decomposed into those weights $\W'$ that hold the critic not smaller than the adaptive control Lyapunov function over the state space.
		A somewhat similar approach was used \eg in \cite{Kim2020}.}
		\blue{For instance, take $V$ as a basis function of the form
			\begin{align*}
				\hat{J}(x,w) = w_1 V(x) + \sum_{j=2}^{l} w_j \varphi_j(x)
			\end{align*}
		to which $w^{\#} = [1 , 0 \dots 0]^\top$ is the recovering weight as per the above assumption. 
		With \eg $\varphi_j(x) \geq V(x)$, $\W' = \{w \in \W: w_1 \geq 1, \; w_j \geq 0, j = 2,\dotsc,l\}$ yields a suitable constraint set}.
		
		Positive definiteness of the approximant can be guaranteed by using \eg \cite[Thm.~2]{Richards2018}, in which the Cauchy-Schwarz inequality and $\W$ ensure an upper bound \blue{and the lower bound is provided by the adaptive control Lyapunov function. 
		That is, 
		\begin{align*}
			q_1(\|x\|) \coloneqq \alpha_1(\|x\|) &\leq V(x) \\
			&\leq \hat{J}(x,w) \leq \overline{w} L_{\varphi}(z,r)\|x\| \eqqcolon q_2(\|x\|).
		\end{align*}}
		 
		\blue{Property ii) is a bounding condition on the gradient and Hessian of the regressor, of which the former is a common assumption (see \eg \cite{Vamvoudakis2014,Bhasin2013}) and the latter is required in establishing a sampling time.} 
	\end{rem}

	\subsection{Algorithm}
	\label{subsec:actor-critic}
	
	
	\blue{To ensure closed loop state convergence, these training rules should satisfy stability conditions. 
	Satisfaction of these conditions is checked at each sampling instance}. 
	\blue{The aim of the algorithm is to ensure practical stability by associating the critic to a given adaptive control Lyapunov function and to establish a variation of $V_c$ of Lemma~\ref{lem:LF-V_c} as effective Lyapunov function for the closed loop.} 
	
	
	
		\blue{Instead of posing constraints on the critic weights $w_k$ and the approximate optimal controller $u_k$, they have to fulfill conditions that may be checked post-computation at each sampling time $k \delta$. 
		If violated, the backup controller $\mu(x_k,\hat{\theta}_k)$ is called and $w_k = w^{\#}$ of Assumption~\ref{asm:struct-equiv} is set. }
	
	
	\blue{In particular, decay along the system is desired. 
		Since only an estimate of $\theta$ is available, it should hold that}
		\begin{align} \label{eq:opt-constr-1}
			\left\langle \frac{\partial \hat{J}}{\partial x}(x_k,w_k), \delta \mathcal{F}(x_k,u_k,\hat{\theta}_k) \right\rangle    \leq - \frac{\delta}{2}\nu(x_{k},\hat{\theta}_k) + \eps_1
		\end{align}
		for some $\eps_1\geq 0$, \emph{along the estimated dynamics under $\hat{\theta}$}.
	
	
	\begin{rem}
		\blue{Condition \eqref{eq:opt-constr-1} may be embedded directly in the optimization and a backup may be called should no admissible solution be found by the optimizer. }
	\end{rem}

	At $k=0$, $\hat{\theta}_0$ is an initial parameter guess.
	The parameter tuning \eqref{eq:param-estim-dynam} is a construct from within the stability proof of Lem.~\ref{lem:LF-V_c} (see also \cite{Krstic1995a}) which using the approximant $\hat{J}$ transfers to 
	\begin{subequations} \label{eq:theta-tuning-sampled}
		\begin{align}
			\hat{\theta}_{k+1} = \; &\hat{\theta}_k + \delta \Gamma \tau(x_k,w_k), \\
			\tau(x,w) \coloneqq &\left( \frac{\partial^\top \hat{J}}{\partial x}(x,w) F(x)  \right)^\top, \label{eq:tau-hatJ}
		\end{align}
	\end{subequations}
	starting at $\hat{\theta}_0$. 
	
	
	
	
	%

	
	The algorithm is summarized in Algorithm~\ref{alg:scheme}, involving several technicalities discussed in the following Section~\ref{sec:main-results}. 
	
	\begin{algorithm}[h]
		\KwIn{pair $(V,\nu)$ as per Assumption~\ref{asm:ACLF-exist},  
			balls $(\ball_R,\ball_r)$, stage cost $r$, initial parameter guess $\hat{\theta}_0$, weigth matrix $\Gamma$, initial state $x_0 \in \ball_{R}$, membership set $\Theta$\;}
		\textbf{Do:}\\
		- select regressor $\varphi$ satisfying Assumption.~\ref{asm:struct-equiv}\;
		- define critic weight sets $\W,\W'$, weight $w^{\#} \in \W$ of Assumption~\ref{asm:struct-equiv} and set $w_0 \in \W'$\;
		- define critic weight update \eqref{eq:critic-opt-practical} or \eqref{eq:GD-update-u-w} with rate $l_w$\;
		- define control input update \eqref{eq:GD-input-params} or \eqref{eq:actor-critic-opt}\;
		- select sampling time $\delta$ as per \eqref{eq:sampling-time-bound}\;
		- determine bounds $\bar{\eps}_{1,w}$ as per \eqref{eq:eps_12-bound}\;
		\While{$k \geq 0$}{
			get $\hat{\theta}_{k+1}$ from \eqref{eq:theta-tuning-sampled}\;
				get $w_k$ from \eqref{eq:critic-opt-practical} or \eqref{eq:GD-update-u-w}\;
				get $u_k$ using \eqref{eq:GD-input-params} or from \eqref{eq:actor-critic-opt}\;
				\eIf{if $(u_k,w_k)$ satisfy i) \eqref{eq:opt-constr-1} with $\eps_1 \leq \bar{\eps}_1$, ii) $\|w_k - w_{k-1}\| \leq \eps_w \leq \bar{\eps}_w$ and iii) $w_k \in \W'$}{
				take $(u_k,w_k)$ as computed\;
				}{set $(u_k,w_k) = (\mu(x_k,\hat{\theta}_k),w^{\#})$\;
				}

			$k \mapsto k+1$\;
		}
		\caption{Algorithm of reinforcement learning with closed-loop stability guarantee based on adaptive control.}
		\label{alg:scheme}
	\end{algorithm}
	

	\section{Main Results}
	\label{sec:main-results}
	
	In the following, practical stability under the resulting control sequence $\{u_k\}_{k \in \N_0}$ of Algorithm~\ref{alg:scheme} in the sense of Definition~\ref{def:pract-stab} is addressed. 
	For certain practical applications, only a subset of the state needs to be practically stabilized which is addressed via an ISS property of the state and modified properties of the adaptive control Lyapunov function and of $\hat{J}$. 
	First, some technicalities are inspected. 
	
	\subsection{Technical preliminaries}
	\label{subsec:technicalities}
	Due to the SH nature and given a sampling period $\delta$, feasibility of \eqref{eq:actor-critic-opt} cannot be ensured inside a small enough vicinity of the origin, in general. 
	Let $\ball_{r^\ast} $ denote a \emph{core ball} for some $0<r^\ast<r$, which can be used to tackle the aforementioned. 
	That is, once $x_k \in \ball_{r^\ast}$, the setting $\{u_k,w_k\} = \{\mu(x_k,\hat{\theta}_k),w^{\#}\}$ is used, where $w^{\#}$ is the adaptive control Lyapunov function recovering weight of Assumption~\ref{asm:struct-equiv}. 
	
	
	\blue{The following Lipschitz condition is satisfied for the system $\F$ due to differentiability: for any $0< r < \infty$, any compact $\Theta \subset \R^p, \U \subset \R^m$ and any $z \in \R^n$, $\exists L_{\mathcal{F}}(z,r)>0$ such that for all $x,y \in \ball_r(z), \theta \in \Theta, u \in \U$}
	\begin{align} \label{eq:system-Lipschitz}
		\blue{\|\mathcal{F}(x,u,\theta) - \mathcal{F}(y,u,\theta)\| \leq L_{\mathcal{F}}(z,r) \|x-y\|.}
	\end{align}
	
	Let
	\begin{align} \label{eq:bar-J}
		\bar{J} \coloneqq \max_{\subalign{ x &\in \ball_R \\ w &\in \W }} \; \hat{J}(x,w).
	\end{align}

	\blue{Motivated by $V_c$ of Lemma~\ref{lem:LF-V_c}, a stability guarantee is established in the subsequent Section~\ref{sec:main-results} using the Lyapunov function candidate 
		\begin{align}
			\V_c(x, \tilde{\theta}) \coloneqq \hat{J}(x,w) + \frac{1}{2} (\theta - \hat{\theta})^\top \Gamma^{-1} (\theta - \hat{\theta}).
		\end{align}
		Observe that with Assumption~\ref{asm:struct-equiv}, there exists $q_{1,2,3,4} \in \K_{\infty}$ such that
		\begin{align}
			q_1(\|x\|) + q_3(\|\tilde{\theta}\|) \leq \V_c(x,\tilde{\theta}) \leq q_2(\|x\|) + q_4(\|\tilde{\theta}\|),
		\end{align}
		in which $q_3(\|\tilde{\theta}\|) = \tfrac{1}{2}\lambda_{\text{min}}(\Gamma^{-1}) \|\tilde{\theta}\|^2$ and $q_4(\|\tilde{\theta}\|) = \tfrac{1}{2}\lambda_{\text{max}}(\Gamma^{-1}) \|\tilde{\theta}\|^2$.
		Regarding the overshoot and core balls as well as respective state membership, note that
		\begin{align}
			\label{eq:state-membership-inside}
			q_1(\|x\|) \leq \hat{J}(x,w) \leq q_1(r) \; \; \Ra \; \; &x \in \ball_{r}
			\intertext{for any $r\geq 0$ and with $r^\ast \coloneqq q_2^{-1}(v^\ast/2)$,}
			\label{eq:state-membership-outside}
			\frac{v^\ast}{2} <  \hat{J}(x,w) \leq q_2(\|x\|) \; \; \Ra \; \; &x \notin \ball_{r^\ast}
		\end{align}
		for any $v^\ast >0$. 
		Ensuring that the core ball radius satisfies $r^\ast < r$, take $v^\ast \coloneqq q_1(r)$. 
	}
	
	
	
	\subsection{Practical Stability}
	\label{subsec:stability}
	
	Now, practical stability is discussed. 
	\blue{In particular, bounds of $\eps_{1,w}$ as well as $\delta$ are established to ensure decay of a Lyapunov function candidate.} 
	
	\begin{thm} \label{thm:stability}
		Consider system \eqref{eq:sys} in the SH-mode \eqref{eq:sys-SH}. 
		Let Assumptions \ref{asm:ACLF-exist} and \ref{asm:struct-equiv} hold. 
		Assume that $\theta \in \Theta$, where $\Theta \subset \R^p$ is compact. 
		Then there exists $ \bar{\delta} >0$ and $\bar{\eps}_{1,w} \geq 0$ such that if the sampling time satisfies $0<\delta \leq \bar{\delta}$, then $\{u_k\}_{k \in \N_0}$ of Algorithm~\ref{alg:scheme} practically semi-globally stabilizes \eqref{eq:sys} as per Definition~\ref{def:pract-stab}.
	\end{thm}
	
	The proof, which can be found in the appendix, is organized as follows: 
	First, some technicalities are reviewed and a Lyapunov function candidate is proposed. 
	Then, conditions are established such that the trajectory remains inside $\ball_{R^{\ast}}$ as well as $\ball_{r}$ after some (computable) time. 
	Finally, decay of the candidate is established by constraining $\eps_{1,w}$ and $\delta$ suitably. 
	
	In practice, sometimes it is not required to steer the entire state vector to the target but only certain states $x_{\{I\}}$ of cardinality $1 \leq d < n$ while the other system states remain bounded. 
	Such situation arises when a particular subsystem of \eqref{eq:sys} \ie a subset of states $x_{\{n-d\}}$, is input-to-state (ISS) stable \cite{Sontag1990} \wrt to the states $x_{\{I\}}$.  
	The system in Section~\ref{subsec:car-traction-control-study} inherits such property. 
	\begin{asm} \label{asm:ISS-state-subsys-req}
		Given any $I \subset \{1,\dotsc,n\}$, let Assumption~\ref{asm:ACLF-exist} hold for a \pd $\nu:\R^{I} \times \R^p \ra \R_{\geq 0}$ and a \pd $\alpha_{\nu}:\R^{I} \ra \R_{\geq 0}$ \ie the decay of $V$ is dependent on $x_{\{I\}}$. 
		Additionally, there exists a partition $V_{\{I\}}:\R^I \ra \R_{\geq 0}$ of $V$ and $\hat{J}_{\{I\}}:\R^I \times \R^l \ra \R_{\geq 0}$ to there is a recovering weight $w$ as in Assumption~\ref{asm:struct-equiv} for all $x_{\{I\}}$ and there exist bounding functions $q_1,q_2$ such that $q_1(\|x_{\{I\}}\|) \leq \hat{J}(x_{\{I\}},w) \leq q_2(\|x_{\{I\}}\|)$. 
	\end{asm}
	Under these modifications, the following relation to the preceding theorem can be made:
	\begin{prp} \label{prp:stability-subsys}
		Let $I \subset \{1,\dotsc,n\}$, $\bar{d} = n-I \in [0,n]$ and $x_{\{I\}}$ be the state to an $I$-dimensional subsystem of \eqref{eq:sys} such that $x$ can be written as $x = [x_{\{I\}} \; x_{\{\bar{d}\}}]^\top$. 
		Let Assumption \ref{asm:ACLF-exist}--\ref{asm:ISS-state-subsys-req} hold and $\bar{d}>0$. 
		If the $\{\bar{d}\}$-subsystem is ISS \wrt to the $\{I\}$-subsystem, then Theorem~\ref{thm:stability} applies to the $\{I\}$-subsystem \ie the set $\underline{\ball}_r := \{x \in \R^n: \|x_{\{I\}}\| \leq r \} \subset \R^{n - \bar{d}}$ is practically stabilized, while $x_{\{\bar{d}\}}$ remains bounded.
	\end{prp}
	
	
%
%

	
	\section{Case Study}
	\label{sec:case-study}
	
	\blue{As remarked in Section~\ref{sec:preliminaries}, finding an adaptive control Lyapunov function -- in particular satisfying Assumption \ref{asm:ACLF-exist} -- is potentially difficult. 
		Subsequently the approach is more suitable for lower dimensional systems. 
		However, dynamics of special form allow the recursive construction of (adaptive) control Lyapunov functions efficiently as \eg in the case of systems in (parametric) strict-feedback form \cite[Thm.~3.5]{Krstic1995}}
	\begin{align*}
		\dot{x}_1 &= x_2 + F_1^\top(x_1) \theta \\
		\dot{x}_2 &= x_3 + F_2^\top(x_1,x_2) \theta \\
		&\hspace*{6pt}\vdots \\
		\dot{x}_{n-1} &= x_n + F_{n-1}^\top(x_1,\dotsc,x_{n-1}) \theta \\
		\dot{x}_n &= F_n^\top(x) \theta + g(x) u, \qquad g(x) \neq 0, \, \forall x. 
	\end{align*}
	
	Since the state of the presented practically stabilizing reinforcement learning controller must not, in general, converge to the origin, the infinite-horizon cost \eqref{eq:IH-cost} may not be finite. 
	Subsequently, another measure of the controller's performance should be considered.  
	In the comparison of the learning controller to the adaptive control Lyapunov function based controller, only the transient performance until the target set is reached after some time $K \in \N_0$ is considered \eg by
	\begin{align} \label{eq:exm-IH-cost-compar}
	\mathcal{C}_{\%}(x_0) =  \frac{\sum_{k=0}^{K_u} r(x_k,u_k)}{\sum_{k=0}^{K_{\mu}}r(x_k,\mu(x_k,\hat{\theta}_k)} \, \text{d}t, 
	\end{align}
	where $K_{u,\mu} \in \N_0$ denote the respective times required until the state reaches $\ball_{r}$ from $x_0 \in \ball_{R}$.

	\blue{To highlight the potentials of a known adaptive control Lyapunov function, consider the following two case studies.} 
	
	\subsection{Car Traction Control}
	\label{subsec:car-traction-control-study}
	
\begin{figure}[h]
		\centering\includegraphics[scale = 0.6]{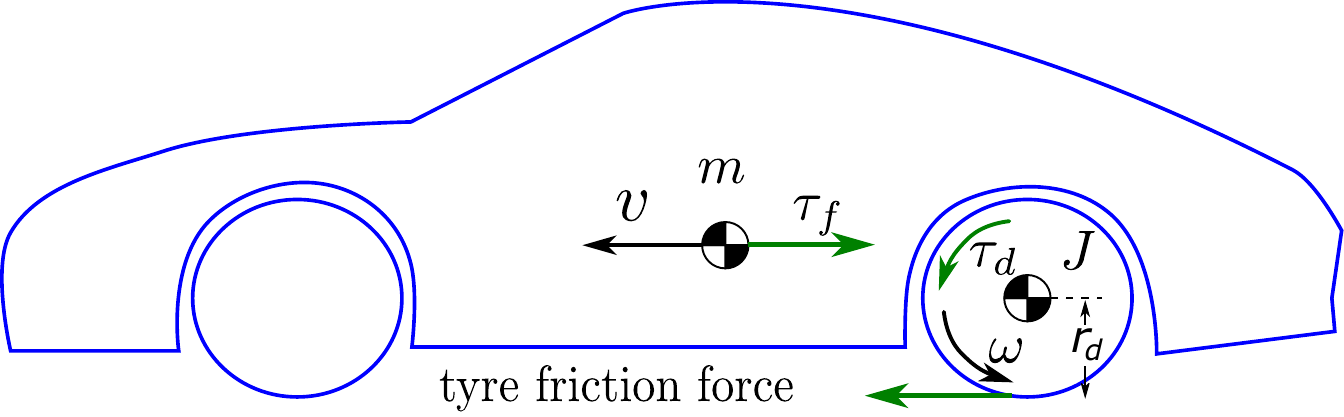}
		\caption{Model of the vehicle and its state and parameters. 
		}
		\label{fig:car}
	\end{figure}

	
	Consider the following dynamical traction model 
	\begin{align} \label{eq:exm-car-dyn}
		\begin{split}
			\dot{v} &= \frac{1}{m} \left( mg \lambda(s) - \tau_f \right) \\
			\dot{s} &= a_1 \frac{\omega}{v} + \frac{mg}{v}h(s)\lambda(s)-a_2\frac{1}{v}\tau_d-a_3\frac{\omega}{v^2}\tau_f
		\end{split}
	\end{align}
	with a slip ratio $s$, horizontal speed $v$, control input $\tau_d$ and
	\begin{align}
		\lambda(s) &= c_1\left(1-e^{-c_2s} \right) - c_3 s + \Delta\lambda(s)  \label{eq:lambda-of-s} \\
		h(s) &= \frac{1}{m}\left( 1- s \right) + \frac{r^2}{J}.  \label{eq:h-of-s}
	\end{align}
	Here, $\Delta\lambda(s)$ denotes a (Lipschitz continuous) uncertainty and $a_{1,2,3}>0$ as well as $c_{1,2,3}>0$ are constants (values of the latter depending on the road surface). 
	Furthermore, $m,g,\lambda(s),J,\tau_f,\omega,r$ denote the vehicle mass, the gravitational acceleration, the friction coefficient, the wheel inertia, a condensed friction term referred to as braking torque, the angular speed of the wheel and its radius, respectively.  
	A schematic of the model is presented in Figure \ref{fig:car}. 
	The goal is to control the slip ratio $s$ to a constant reference $s^\ast \in (0,1)$ -- the specific task herein is to stabilize $s^\ast = 0.2$. 
	
	The given system falls in the category of Proposition~\ref{prp:stability-subsys}. 
	Specifically, the $v$-coordinate can be made input-to-state stable (ISS) with respect to the input $s$ given a specific control $\tau_d$ -- the reader may refer to \cite{Nakakuki2008} for a detailed derivation. 
	For brevity, denote $x:= [v,\, s]^\top$. 
	It can be shown that the function 
	\begin{align} \label{eq:exm-car-ACLF}
		V(x,\theta) = V(x) = \frac{1}{2} \tilde{x}_2^2,
	\end{align}
	where $\tilde{x}_2 := x_2 - x_2^\ast$ and $x_2^\ast = s^\ast$ is the slip ratio set point, is an adaptive control Lyapunov function for \eqref{eq:exm-car-dyn} that satisfies Assumption~\ref{asm:ACLF-exist}. 
	In the given setting, $\theta = \begin{bmatrix}	\varrho & c_3 & M & \tau_f \end{bmatrix}^\top$ represent uncertain parameters of the system as well as of ISS specific bounds.
		The coefficients $c_1 = 1.2801$, $c_2 = 23.99$ and $c_3 = 0.52$ in \eqref{eq:lambda-of-s} represent dry asphalt conditions and $\Gamma,\eps,k,M,J,r,d$ are set according to \cite{Nakakuki2008}.  
	
	According to \cite{Nakakuki2008}, given the adaptive control Lyapunov function \eqref{eq:exm-car-ACLF}, the controller $\tau_d = \mu$ reads as
	\begin{align} \label{eq:exm-car-ctrl}
		\begin{split}
			\mu(x,\hat{\theta}) = &\frac{a_1}{a_2}\omega + \left( \frac{x_1}{a_2}k + \frac{mg}{a_2}h(x_2)\hat{M} \right) \tilde{x}_2 \\
			&- \frac{a_3 \omega}{a_2 x_1}\hat{\tau}_f + \frac{mg}{a_2}h(x_2) \left( \hat{\varrho}\left(1-e^{-\varkappa x_2} \right) - \hat{c}_3 x_2 \right),
		\end{split}
	\end{align}
	with $\hat{\theta} = \begin{bmatrix}	\hat{\varrho} & \hat{c}_3 & \hat{M} & \hat{\tau}_f \end{bmatrix}^\top$ and a controller gain $k>0$, and the parameter estimator dynamics \eqref{eq:param-estim-dynam} are given through
	\begin{align} \label{eq:exm-car-tau}
		\begin{split}
			&\tau(x,\theta) = \tau(x) = \Phi(x) \tilde{x}_2, \\
			&\Scale[0.85]{\Phi := \begin{bmatrix}
					-\frac{a_3 \omega}{x_1^2} & \frac{mg}{x_1}h(x_2)(1-e^{-\varkappa x_2}) & -\frac{mg}{x_1}h(x_2)x_2 & \frac{mg}{x_1}h(x_2)\tilde{x}_2
				\end{bmatrix}^\top}.
		\end{split}
	\end{align}
	Then, a Lyapunov function for system \eqref{eq:exm-car-dyn}--\eqref{eq:h-of-s} with \eqref{eq:exm-car-ctrl} and \eqref{eq:exm-car-tau} reads as
	\begin{align}
		V_c(x,\tilde{\theta}) = \frac{1}{2} \tilde{x}_2^2 + \frac{1}{2} \tilde{\theta}^{\top} \Gamma^{-1} \tilde{\theta},
	\end{align}
	which attains a decay rate of $\dot{V}_c =  k\tilde{x}_2^2$. 
	
	Note that the decay rate complies with Assumption~\ref{asm:ACLF-exist} as it can be lower bounded uniformly.
	
	Regarding Assumption~\ref{asm:struct-equiv} and \ref{asm:ISS-state-subsys-req}, take $\varphi = \tilde{x}_2^2$ for simplicity. 
	Hence $w^{\#} = 0.5$ and \blue{$\W'$ consist of values $w_k \geq w^{\#}$. 
	Suitable weight norm bounds for $\W$ can be selected arbitrarily. }
	Computation of $w_k$ is done via \eqref{eq:critic-opt-practical} for $x^s = x_k$ only. 
	The stage cost is chosen as 
	\begin{align*}
		r(x,u) = 5\tilde{x}_2^2  + 0.1 u^2. 
	\end{align*}
The sampling time is set to $\delta = 0.01$. 
	The parameter estimate $\hat{\theta}$ is initialized with zeros in all entries \ie $\hat{\theta}_0 =0$.  
	
\begin{figure}[h]
		\centering
		\begin{subfigure}[t]{0.5\textwidth}
			\centering
			\includegraphics[width=0.7\columnwidth]{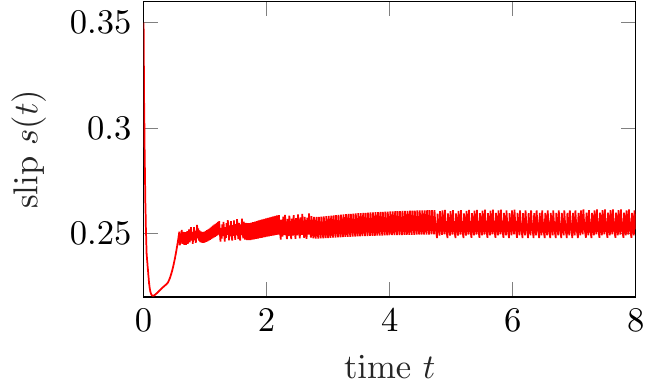}
		\end{subfigure}%
	\hfill
	\begin{subfigure}[t]{0.5\textwidth}
		\centering
		\includegraphics[width=0.7\columnwidth]{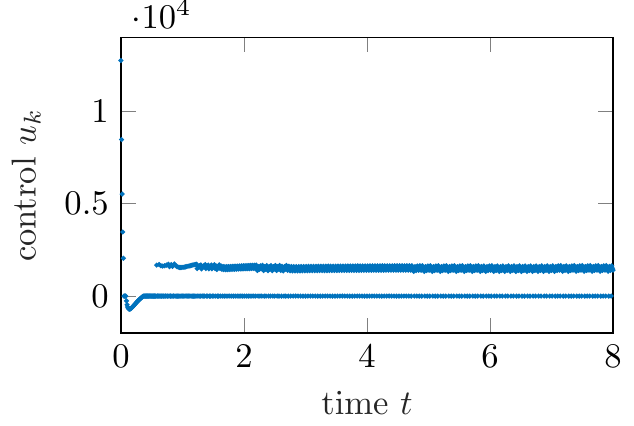}
	\end{subfigure}
\caption{Practically stable slip and associated control signal for $x_0 = [85 \; \; 0.35]^\top$. To keep the state in a vicinity of the reference slip, the input is required to chatter.}
\label{fig:car_traction_slip_input}
	\end{figure}

	Fig.~\ref{fig:car_traction_slip_input} depicts the tire slip of the car, converging to the target ball, and the associated control signals. 
	As is common for sampled control Lyapunov function based stabilization (refer to \eg \cite{Braun2017}), the input experiences chattering behavior. 

\begin{figure}[H]
		\centering
		\includegraphics[width=0.4\textwidth]{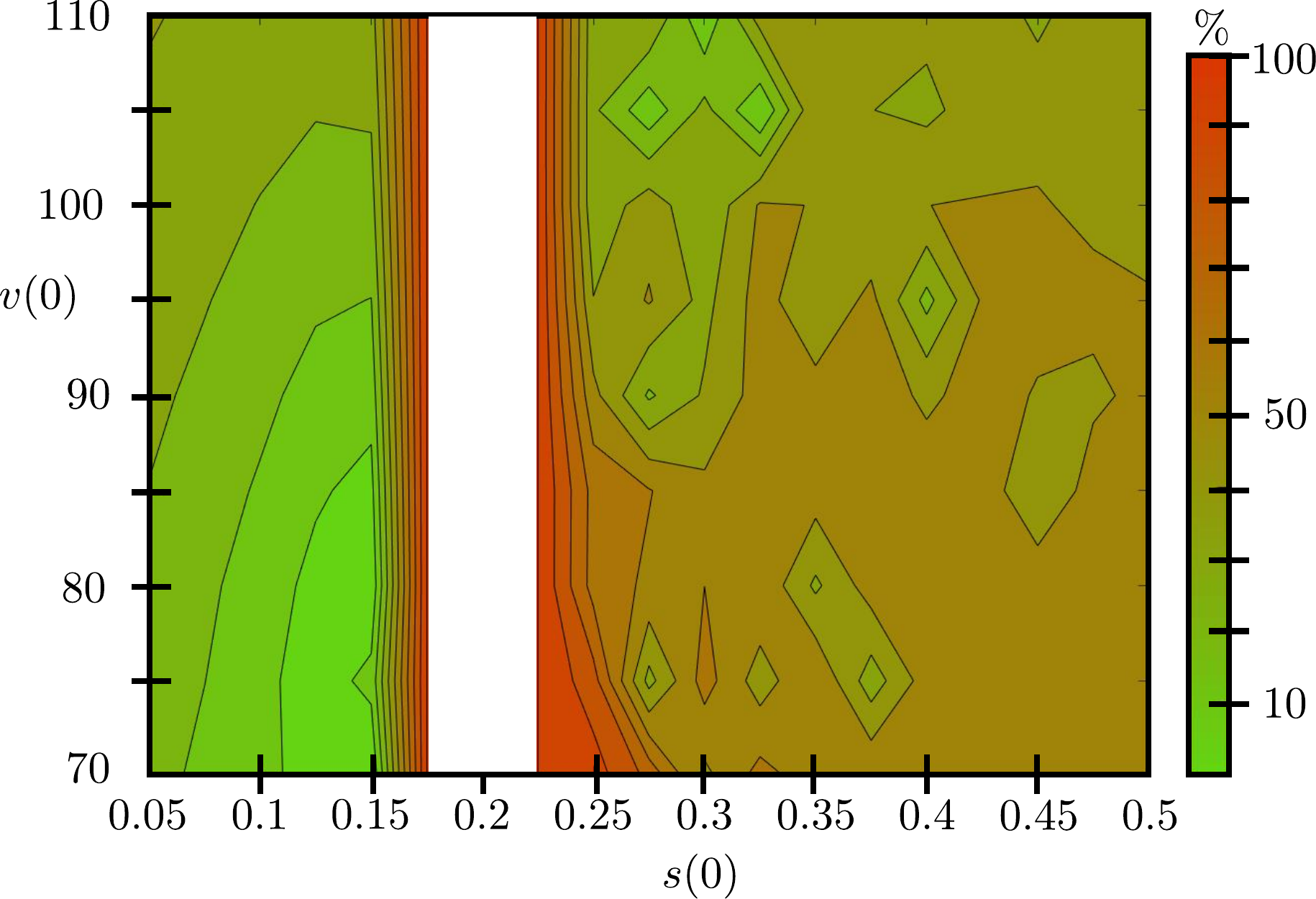}
		\caption{Contour of $\mathcal{C}_{\%}(x_0)$ over a grid $x_0 \in [70,110] \times [0.05,0.5]$. A cost reduction of up to $90 \%$ can be achieved compared to the adaptive controller $\mu$. A possible conclusion may be that if the initial state is near the target set, the optimizer tends to resorts to the feasible action $\mu$ and thereby achieves less performance improvement.}
		\label{fig:contour-cost}
	\end{figure}

	For a selected range of initial states $x(0) = [v(0), \, s(0)]^\top$, the cost difference $\mathcal{C}_{\%}(x_0)$ for reaching the target set $\| s - s^\ast \| \leq r = 0.03$ under the control sequence $\{u_k\}_{k \in \N_0}$ and the adaptive controller $\{\mu(x_k,\hat{\theta}_k)\}_{k \in \N_0}$ is shown in Fig.~\ref{fig:contour-cost}. 
	It can be seen that the reinforcement learning scheme outperforms the baseline controller $\mu$. 
	An increased computational effort may be expected when $u_k$ is obtained via \eqref{eq:actor-critic-opt} -- \blue{this may be circumvented, however, by resorting to gradient based rules as per \eqref{eq:GD-input-params}}. 

	\blue{Though the cost comparison reveals significant performance improvement, it should be kept in mind that factors such as \eg the critic update rule, sampling frequency or the initial parameter $\hat{\theta}_0$ may have influence on the performance. 
	} 
	

%
%
	Aside confirming the functionality of the proposed control scheme of Section~\ref{subsec:actor-critic} \ie practical stability by a reinforcement learning controller, another point of view of this work is to additionally reveal the potentials of ``augmenting'' the baseline adaptive controller with an ADP strategy.

	\subsection{\blue{Adaptive Cruise Control}}
	\label{subsec:cruise-control}
	
	As a second example, consider a simplified adaptive cruise control problem as investigated in \cite{Ames2014,Taylor2020}. 
	The system dynamics are given by
	\begin{align}
	\begin{split}
		\dot{v} = &-\frac{1}{m}\left( f_0  + v \, f_1 + v^2 \, f_2  \right) + \frac{1}{m} u 
		\end{split}
	\end{align}
	with $v$ describing the velocity of a vehicle. 
	The task is bring the velocity to a desired value $v^\ast$ while $f_{0,1,2}$ are unknown frictional force parameters. 
	Suppose henceforth $v^\ast= 14$. 
	These parameters will be estimated by \eqref{eq:tau-hatJ} to yield $\hat{\theta} = [\hat{f}_0 \; \hat{f}_1 \; \hat{f}_2]^\top$. 
	It was shown in \cite{Ames2014} that the function
	\begin{align}
		V(v,\theta) = V(v) = \tilde{v}^2 
	\end{align}
	with $\tilde{v} = v -v^\ast$ is an adaptive control Lyapunov function associated with the controller
	\begin{align}
		\mu(v,\hat{\theta}) = - \eps \frac{m}{2} \tilde{v} + \hat{f}_0 + \hat{f}_1 v + \hat{f}_2 v^2
	\end{align}
	for some $\eps>0$. 
	The effective Lyapunov function $V_c$ is thus given by
	\begin{align}
		V_c(x,\tilde{\theta}) = \tilde{v}^2  + \frac{1}{2} \tilde{\theta}^\top \Gamma^{-1} \tilde{\theta}.
	\end{align}
	Again, for simplicity, take $\varphi = [\tilde{v}^2]^\top$ such that $w^{\#} = 1$ and $V(x) \leq \hat{J}(x,w)$ for $w \geq 1$. 
	All system and control associated parameters can be found in \cite{Ames2014}. 
	Therein, it was observed that the stabilizing input $\mu$ may take high values.  
	To comply with the desirable bound $\eps_w$ as to ensure decay of $\V_c$ (refer to Theorem~\ref{thm:stability}), the learning rate $l_w$ and the weight matrix $R$ of the stage cost $r(x,u) = x^\top Q x + u^\top R u$ with 
	\begin{align*}
		r(v,u) = \tilde{v}^2 + 10^{-7} u^2
	\end{align*}
	are chosen sufficiently small. 
	Note that by choosing a small learning rate, switches the the backup controller may occur less frequent. 
	Note further that $R$ can be chosen freely and the condition can be met by the choice of $l_w$. 
	The sampling time is set to $\delta = 0.01$ and $w_0 = w^{\#}$. 
	
	\begin{figure}[h]
		\centering
		\begin{subfigure}[t]{.5\textwidth}
			\centering
			\includegraphics[width = 0.7\columnwidth]{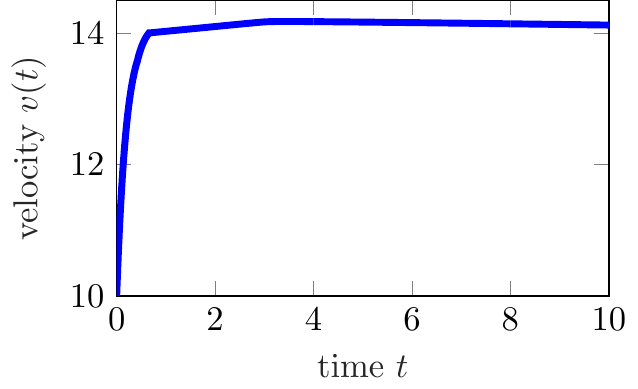}
		\end{subfigure}%
	\hfill
	\begin{subfigure}[t]{.5\textwidth}
		\centering
		\includegraphics[width = 0.7\columnwidth]{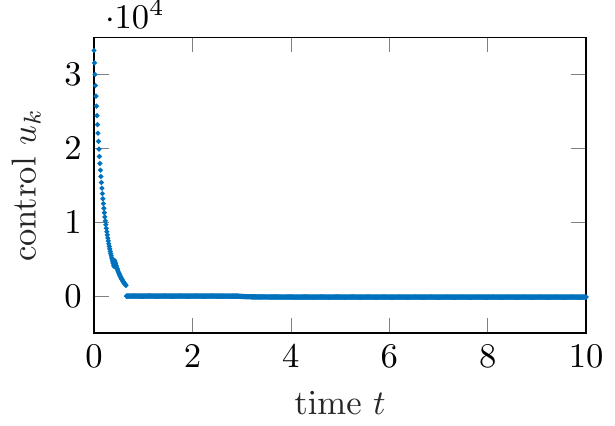}
	\end{subfigure}%
\hfill
\begin{subfigure}[t]{.5\textwidth}
	\centering
	\includegraphics[width = 0.7\columnwidth]{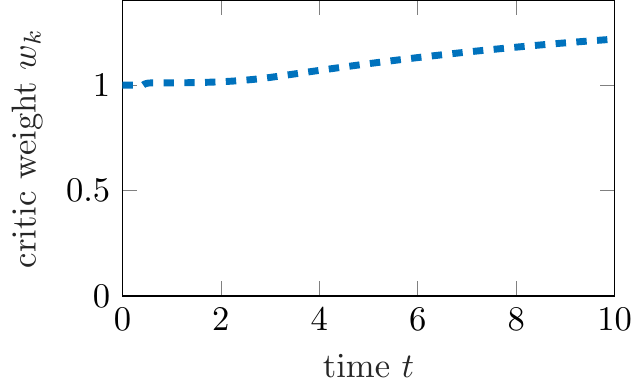}
\end{subfigure}%
\hfill
\begin{subfigure}[t]{.5\textwidth}
	\centering
	\includegraphics[width = 0.7\columnwidth]{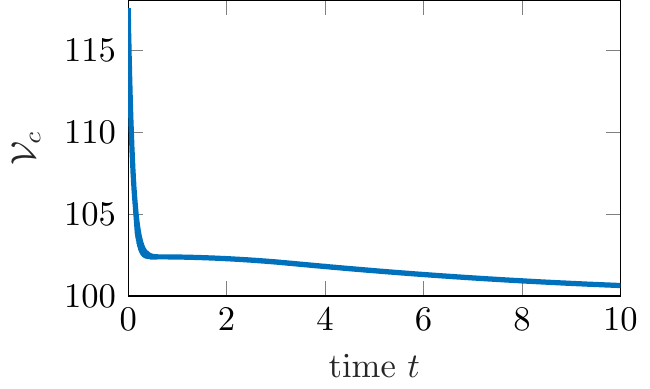}
\end{subfigure}%
\hfill
\caption{Practically stable velocity of the vehicle. The weight $w_k$ of the regressor changes sufficiently small so that decay of $V_c$ is given as the input $u_k$ satisfies the decay condition. No backup $(\mu(x_k,\hat{\theta}_k),w^{\#})$ is required.}
\label{fig:cruise_control}
	\end{figure}

Fig.~\ref{fig:cruise_control} depicts the practically stable state with respect to $v^\ast$ and the associated control values $u_k$ together with the weight $w_k$ of the critic and the effective Lyapunov function of the system $\V_c$. 
It was observed that $\mathcal{C}_{\%}(v_0 = 10) \approx 0.12$ which indicates a significant performance improvement. 
	
	
	\section{Conclusion}
	\label{sec:conclusion}
	
	
	This work presented a reinforcement learning method with a closed-loop practical stability guarantee.
	The stability guarantee was achieved vie employing techniques of adaptive control. 
	By means of a sample-and-hold control, it was shown that the system state converge to a prescribed vicinity of the origin, that does not depend on the approximation quality of the optimal cost function or ``richness'' of the respective approximating structure. 
	A set of conditions for the reinforcement learning controller was presented so as to ensure closed-loop stability. 
	If a condition is violated, the backup adaptive control Lyapunov function and its associated stabilizing controller are invoked. 
	Knowledge of the adaptive control Lyapunov function is a particular issue postponed to future studies but which may involve the use of deep neural net structures in the critic. 
	
	\appendix
	
	\begin{IEEEproof}(of Thm. \ref{thm:stability}) 
		Consider the Lyapunov function candidate
		\begin{align*}
			\V_c(x,\tilde{\theta}) = \hat{J}(x,w) + \underbrace{\frac{1}{2} (\theta - \hat{\theta})^\top \Gamma^{-1} (\theta - \hat{\theta})}_{\eqqcolon \T(\tilde{\theta})}
		\end{align*}
		and define its difference
		\begin{align*}
			\Delta_k \V_c = \Delta_k \hat{J} + \Delta_k \T
		\end{align*}
		along \eqref{eq:sys-SH} and \eqref{eq:theta-tuning-sampled} with
		\begin{align*}
			\Delta_k \hat{J} &\coloneqq  \hat{J}(x_{k+1}(u_k),w_{k+1}) - \hat{J}(x_k,w_k) \\
			\Delta_k \T &\coloneqq \T(\tilde{\theta}_{k+1}) - \T(\tilde{\theta}_k)
		\end{align*} 
		and $\tilde{\theta}_k = \theta - \hat{\theta}_k$. 
		
		First, boundedness of the state trajectory is addressed. 
		Recalling that $\hat{\theta}_k$ is updated in discrete time steps $k \in \N_0$, the Lipschitz property yields
		\begin{align*}
			\hat{J}(x(t;u_k),w_k) + \T(\tilde{\theta}_k) \leq \hat{J}(x_k,w_k) + \T(\tilde{\theta}_k) + \overline{w} L_\varphi \delta \bar{\F}
		\end{align*}
		for $t \in [0,\delta)$, with $L_{\varphi}$ the Lipschitz constant of $\varphi$ on a to be defined domain.  
		Assuming that $\hat{J}(x_k,w_k) \leq \bar{J}$, specifying $R^\ast>R$ gives the respective $\delta>0$ such that 
		\begin{align} \label{eq:condition-overshoot-ball}
			\bar{J} + \overline{w} L_\varphi \delta \bar{\F} \leq q_1(R^\ast)
		\end{align}
		with $L_{\varphi} = L_{\varphi}(0,R^\ast)$, implying that
		\begin{align} \label{eq:V_c-SH-traj}
			\begin{split}
				\V_c(x(t;x_k),\tilde{\theta}_k) \leq &\hat{J}(x_k,w_k) + \T(\hat{\theta}_k) + \overline{w} L_\varphi \delta \bar{\F} \\
				\leq &\bar{J} + \T(\hat{\theta}_k) + \overline{w} L_\varphi \delta \bar{\F} \\
				\leq &q_1(R^\ast) + q_4(\|\tilde{\theta}_k\|) \\
				\stackrel{-q_4(\|\tilde{\theta}_k\|)}{\Ra} \; \;  \hat{J}(x(t;x_k),w_k) \leq & q_1(R^\ast)
			\end{split}
		\end{align}
		at any $k \in \N_0$ and for any $t \in [0,\delta)$ which in turn gives $x(t;x_k) \in \ball_{R^{\ast}}$ due to \eqref{eq:state-membership-inside}. 
		Since $x_0 \in \ball_{R}$ and $w_0 \in \W$ by constraint, $\hat{J}(x_k,w_k) \leq \bar{J}$ holds at $k =0$ and if $\Delta_k \V_c=	\V_c(x_{k+1},\tilde{\theta}_{k+1})-\V_c(x_k,\tilde{\theta}_k)<0$, the state is bounded to $\ball_{R^{\ast}}$ for all $k \in \N_0$ and inter sampling times $t \in [0,\delta)$. 
		Denote $\U = \U_{\ball_{R^{\ast}}}$ the control set associated to $\ball_{R^{\ast}}$ as per Definition~\ref{def:ACLF}. 
		
		Next, the decay of $\V_c$ is discussed. 
		Consider the following variant of \eqref{eq:state-trajectory-explicit}
		\begin{align*}
			x(t;u) = x_k +\delta\; \underbrace{\left( \, \frac{1}{\delta} \int_{0}^{t} \mathcal{F}(x(k\delta + s),u_k,\theta) \; \text{d}s \, \right)}_{\eqqcolon \Psi_t}
		\end{align*}
		with $t \in [0,\delta)$. 
		Assume for now that the state remains in a ball $\ball_{R^{\ast}}$ with $R^\ast >R$ for all times \ie $x(t;u_k) \in \ball_{R^{\ast}}$, for all $t \in [0,\delta)$ and $k \in \N_0$, which will be shown to hold using $\V_c$ and a suitable choice of $\delta>0$. 
		Let
		\begin{align*}
			\bar{\F} \coloneqq \sup_{\subalign{x \in \ball_{R^{\ast}}&, \; u \in \U \\ \theta \in& \Theta  }} \; \F(x,u,\theta) 
		\end{align*}
		with $\U \coloneqq \U_{\ball_{R^{\ast}}}$ the respective control set associated with the adaptive control Lyapunov function as per Definition~\ref{def:ACLF} and
		\begin{align*}
			\bar{\tau} \coloneqq \sup_{\subalign{x &\in \ball_{R^{\ast}} \\ w &\in \W}} \; \tau(x,w).
		\end{align*} 
		Regarding $\Psi_t$, note that 
		\begin{align*}
			\Psi_t = &\frac{1}{\delta}\int_{0}^{t} \mathcal{F}(x(k\delta + s),u_k,\theta) \, \diff s \\
			= &\frac{1}{\delta} \int_{0}^{t} \F(x(k\delta + s),u_k,\theta) \, \diff s + \frac{t}{\delta} \F(x_k,u_k,\theta) \\
			&- \frac{t}{\delta} \F(x_k,u_k,\theta) \\
			= & \frac{1}{\delta} \int_{0}^{t} \F(x(k\delta + s),u_k,\theta)  -  \F(x_k,u_k,\theta)\, \diff s \\
			&+ \frac{t}{\delta} \underbrace{\F(x_k,u_k,\theta)}_{= \F(x_k,u_k,\hat{\theta}_k) + F(x_k)\tilde{\theta}_k}
		\end{align*}
		as well as $\|\Psi_t\| \leq \tfrac{t}{\delta}\bar{\F} \leq \bar{\F}$ which leads to $\|x(t;u_k) - x_k\| \leq \delta \bar{\F}$.
		By \eqref{eq:system-Lipschitz}, with $L_{\F} \coloneqq L_{\F}(0,R^\ast)$, 
		\begin{align*}
			&\| \int_{0}^t \F(x(k\delta + s),u_k,\theta)  -  \F(x_k,u_k,\theta) \, \diff s\| \\
			\leq &\int_{0}^t L_{\F}\|x(k\delta + s) - x_k\| \, \diff s \leq \delta^2 L_{\F} \bar{\F}.
		\end{align*}

		To tackle $\Delta_k \hat{J}$ in $\Delta_k \V_c$, the Taylor expansions is considered yielding
		\begin{align} \label{eq:Delta-Jhat}
			\begin{split}
				\hat{J}(x_k + \delta \Psi_t,w_k) \leq \; &\hat{J}(x_k,w_k) + \langle \nabla_x \hat{J}(x_k,w_k),\delta \Psi_t \rangle \\
				&+\frac{\delta^2}{2} \underbrace{\|\Psi_t\|^2}_{\leq \F^2} \underbrace{\sup_{\subalign{x &\in \ball_{R^{\ast}} \\ w &\in \W}} \|\nabla_x^2 \hat{J}(x,w)\|}_{\leq \varphi_2}
			\end{split}
		\end{align}
		for any $w_k \in \W$, $x_k \in \ball_{R^{\ast}}$ and $t \in [0,\delta)$. 
		
		Additionally, at $k+1$ the critic weights are updated along \eqref{eq:critic-opt-practical} or \eqref{eq:GD-update-u-w} to yield 
		\begin{align*}
			\Delta_k \hat{J} \leq &\hat{J}(x_{k+1}(u_k),w_k) - \hat{J}(x_{k},w_k) \\
			&+ \langle w_k - w_{k+1}, \varphi(x_{k+1}(u_k)) \rangle \\
			\leq &\hat{J}(x_{k+1}(u_k),w_k) - \hat{J}(x_{k},w_k) + \eps_w \underbrace{\sup_{x \in \ball_{R^{\ast}}} \|\varphi(x)\|}_{\eqqcolon \bar{\varphi}}
		\end{align*}
		
		Regarding $\Delta_k \T$, observe that with \eqref{eq:tau-hatJ},
		\begin{align*}
			\Delta_k \T = &\frac{1}{2} \tilde{\theta}_k^\top \Gamma^{-1} \tilde{\theta}_k - \delta \tau(x_k,w_k)^\top \tilde{\theta}_k \\
			&+ \frac{\delta^2}{2} \tau(x_k,w_k)^\top \Gamma \tau(x_k,w_k) - \frac{1}{2} \tilde{\theta}_k^\top \Gamma^{-1} \tilde{\theta}_k \\
			= &- \delta \tau(x_k,w_k)^\top \tilde{\theta}_k + \frac{\delta^2}{2} \tau(x_k,w_k)^\top \Gamma \tau(x_k,w_k) \\
			\leq &- \delta \tau(x_k,w_k)^\top \tilde{\theta}_k  + \frac{\delta^2}{2} \underbrace{\|\tau(x_k,w_k)\|^2}_{\leq \bar{\tau}^2} \|\Gamma\|.
		\end{align*}
		
		Merging $\Delta_k \hat{J}$ and $\Delta_k \T$ yields
		\begin{align} \label{eq:V_c-difference-intermediate}
			\begin{split}
				&\Delta_k \V_c  \\
				=&\Delta_k \hat{J} + \Delta_k \T \\
				\leq &\hat{J}(x_{k+1}(u_k),w_k) - \hat{J}(x_{k},w_k) + \eps_2 + \Delta_k \T \\
				\leq &\langle \nabla_x \hat{J}(x_k,w_k),\delta \Psi_t \rangle 
				+\frac{\delta^2}{2} \F^2 \varphi_2 \\
				&-  \delta \tau(x_k,w_k)^\top \tilde{\theta}_k  + \frac{\delta^2}{2} \bar{\tau}^2 \|\Gamma\| +  \eps_w\bar{\varphi} \\
				= & \langle \nabla_x \hat{J}(x_k,w_k),t \F(x_k,u_k,\hat{\theta}_k) \rangle \\
				&+ \underbrace{\langle \nabla_x \hat{J}(x_k,w_k),t F(x_k)\tilde{\theta}_k \rangle}_{= t \tau(x_k,w_k)^\top \tilde{\theta}_k} \\
				&+ \langle \nabla_x \hat{J}(x_k,w_k), \int_{0}^{t} \hspace*{-5pt} \F(x(k\delta + s),u_k,\theta)  -  \F(x_k,u_k,\theta) \diff s\rangle \\
				&+\frac{\delta^2}{2} \F^2 \varphi_2 - \delta \tau(x_k,w_k)^\top \tilde{\theta}_k   + \frac{\delta^2}{2} \bar{\tau}^2 \|\Gamma\| + \eps_w\bar{\varphi} 
			\end{split}
		\end{align}
		for $t \ra \delta $, in which the cancellation as per \cite{Krstic1995a} can be performed for $\pm \delta  \tau(x_k,w_k)^\top \tilde{\theta}_k$ resulting in
		\begin{align} \label{eq:V_c-difference-complete}
			\begin{split}
				&\Delta_k \V_c \\
				\leq &\langle \nabla_x \hat{J}(x_k,w_k),\delta \F(x_k,u_k,\hat{\theta}_k) \rangle +\frac{\delta^2}{2} \F^2 \varphi_2 + \eps_w\bar{\varphi} \\
				&+\underbrace{\sup_{\subalign{x &\in \ball_{R^{\ast}} \\ w &\in \W}}\|\nabla_x \hat{J}(x_k,w_k)\|}_{\leq \bar{\varphi}_1 } \; \delta^2 L_{\F}\bar{\F}  + \frac{\delta^2}{2} \bar{\tau}^2 \|\Gamma\| \\
				\leq & - \frac{\delta}{2}\nu(x_{k},\hat{\theta}_k) + \eps_1 + \eps_w\bar{\varphi}  \\
				&+ \delta^2 \underbrace{\big( \bar{\varphi}_1 L_{\F} \bar{\F} + \frac{1}{2} \bar{\tau}^2 \|\Gamma\| + \frac{1}{2} \bar{\F}^2 \bar{\varphi}_2\big)}_{ \eqqcolon \bar{\Delta}},
			\end{split}
		\end{align}
		where the last inequality uses constraint \eqref{eq:opt-constr-1}.

		
		Under $\Delta_k \V_c <0 $, which will be encoded in the choice of $\{\delta,\eps_1,\eps_w\}$, the state will enter the target ball $\ball_{r}$. 
		Several cases emerge. 
		
		
		\textit{Case 1 - $\hat{J}(x_k,w_k) \geq \tfrac{v^\ast}{2}$:} Implies $\V_c(x_k,\tilde{\theta}_k) \geq  \tfrac{v^\ast}{2}$ and by \eqref{eq:state-membership-outside} also that $\|x_k\| > r^\ast$. 
		Define 
		\begin{align} \label{eq:min-decay-rate}
			\bar{\nu} \coloneqq \inf_{r^\ast \leq \|x\| \leq R^\ast} \; \frac{1}{2}\alpha_{\nu}(x). 
		\end{align}
		Selecting $\delta>0$ such that 
		\begin{align} 
			\delta^2 \bar{\Delta} \leq \frac{\bar{\nu}}{10} \delta
		\end{align}
		as well as
		\begin{align} \label{eq:eps_12-bound}
			0 \leq \eps_{1} \leq \frac{2 \bar{\nu}}{5} \delta \eqqcolon \overline{\eps}_1, \quad 0 \leq \eps_{w} \leq \bar{\varphi}^{-1} \frac{2\bar{\nu}}{5} \delta \eqqcolon \overline{\eps}_w
		\end{align}
		yields decay $\Delta_k \V_c = -\bar{\nu} \delta / 10 < 0$. 
		
		\textit{Case 1.1 - repeated violation:} If at two consecutive sampling times $k\delta$ and $(k+1)\delta$ the desired decay is not achieved \ie the bounds $\bar{\eps}_{1,w}$ are not satisfied for $(u_k,w_k)$, the backup pair $(\mu(x_k,\hat{\theta}_k),w^{\#})$ will result in recovering the decay of the Lyapunov function $V_c$ associated with $V$ as per Lemma~\ref{lem:LF-V_c} on the set $\ball_{R^{\ast}} \setminus \ball_{r^\ast}$ as well as $\tau$ of the parameter estimation dynamics \eqref{eq:param-estim-dynam} as follows. 
		
		With $\hat{J}(x_k,w_k) = V(x_k)$, $\V_c$ reduces to $V_c$ of Lemma~\ref{lem:LF-V_c} and the decay rate for $\delta \ra 0$ will be given by $\nu$ as per \eqref{eq:decay-V_c}. 
		Since along $\F(x,u,\theta)$ and $\Gamma \tau(x,w^{\#})$, the Taylor expansion yields
		\begin{align*}
			&V_c\left(x_{k+1}(\mu(x_k,\hat{\theta}_k)),\tilde{\theta}_{k+1} \right) - V_c(x_k,\tilde{\theta}_k) \\
			\leq &-\delta \nu(x_k,\hat{\theta}_k) + \sigma(\delta^2) 
		\end{align*}
		for any $\delta>0$, given balls $\ball_{r^\ast,R^\ast}$, there exists $\bar{\delta}_a >0$ such that for any $0 < \delta \leq \bar{\delta}_a$, for all $x_k \in \ball_{R^\ast} \setminus \ball_{r^\ast}$, $k \in \N_0$, 
		\begin{align} \label{eq:decay-V_c-sampled}
			\begin{split}
				\Delta_k \V_c = &V_c\left(x_{k+1}\left(\mu(x_k,\hat{\theta}_k)\right),\tilde{\theta}_{k+1} \right) - V_c(x_{k},\tilde{\theta}_{k}) \\
				\leq &-\frac{\delta}{2} \nu(x_{k},\hat{\theta}_k) \leq -\delta \bar{\nu},
			\end{split}
		\end{align}
		with consecutive employment of $w_{k,k+1} = w^{\#}$ and $u_k = \mu(x_k,\hat{\theta}_k)$. 
		
		\textit{Case 1.2 - single violation:} Consider \eqref{eq:V_c-difference-intermediate} and \eqref{eq:V_c-difference-complete}, by which 
		\begin{align*}
			\Delta_k \V_c = &\Delta_k \hat{J} + \Delta_k \T \\
			= &\hat{J}(x_{k+1}(u_k),w_{k+1})- \hat{J}(x_{k},w_k) + \Delta_k \T \\
			&-\hat{J}(x_{k+1}(u_k),w_k) + \hat{J}(x_{k+1}(u_k),w_k)  \\
			\leq &- \frac{\delta}{2}\nu(x_{k},\hat{\theta}_k) + \eps_1 + \delta^2 \bar{\Delta} \\
			&+ \hat{J}(x_{k+1}(u_k),w_{k+1}) -\hat{J}(x_{k+1}(u_k),w_k). 
		\end{align*}
		If the conditions are violated at $k+1$ but hold at $k$, $w_{k+1} = w^{\#}$ and thus
		\begin{align*}
			&\underbrace{\hat{J}(x_{k+1}(u_k),w_{k+1})}_{= V(x_{k+1}(u_k))} - \underbrace{\hat{J}(x_{k+1}(u_k),w_k)}_{\geq V(x_{k+1}(u_k))} \\
			\leq &V(x_{k+1}(u_k)) - V(x_{k+1}(u_k)) = 0.
		\end{align*}
		If the conditions are violated at $k$ \ie $w_k = w^{\#}$, but can be recovered at $k+1$, then $u_k = \mu(x_k,\hat{\theta}_k)$ and $\hat{J}(x_{k+1}(u_k),w_{k+1}) \leq V(x_{k+1}(u_k)) + \|w_{k+1} - w^{\#}\| \bar{\varphi}$ can be used to obtain 
		\begin{align*}
			\Delta_k \V_c \leq -\delta \bar{\nu} + \|w_{k+1} - w^{\#}\| \bar{\varphi}
		\end{align*}
		from \eqref{eq:decay-V_c-sampled}. 
		Since $\eps_w$ upper bounds $\|w_{k+1} - w_{k}\|$ with margin \eqref{eq:eps_12-bound}, $\Delta_k \V_c \leq - \delta 3\bar{\nu}/5$. 
		
		\textit{Case 2 - $\V_c(x_k,\tilde{\theta}_k) \leq \tfrac{3v^\ast}{4}$:} Due to the specification of $v^\ast$ as in Section~\ref{subsec:actor-critic} and
		\begin{align*}
			q_1(\|x_k\|) \leq \hat{J}(x_k,w_k) \leq \V_c(x_k,\tilde{\theta}_k) \leq \tfrac{3v^\ast}{4} \, \Ra \, \|x_k\| \leq \frac{3 r^\ast}{4} ,
		\end{align*}
		the state has entered the target ball. 
		Using \eqref{eq:V_c-SH-traj}, if $\delta>0$ satisfies
		\begin{align*}
			\bar{w} L_{\varphi} \delta \bar{\F} \leq \frac{v^\ast}{4}
		\end{align*}
		then
		\begin{align*}
			\V_{c}(x(t;x_k),\tilde{\theta}_k) \leq  \underbrace{\V_{c}(x_k,\tilde{\theta}_k)}_{\leq \frac{3v^\ast}{4}} + \underbrace{\bar{w} L_{\varphi} \delta \bar{\F}}_{\leq \frac{v^\ast}{4}} \leq v^\ast.
		\end{align*}
		Subsequently $\|x(t;x_k)\| \leq r$, $t\in [0,\delta)$. 
		Being in Case 1, the state may either remain there or trigger Case 2.

		Therefore, choosing the sampling time $0< \delta \leq \bar{\delta}$, where
		\begin{align} \label{eq:sampling-time-bound}
			\begin{split}
				\bar{\delta} \coloneqq \max_{0< \delta \leq \delta_a} \; &\delta \\
				\text{such that} \quad &\bar{w} L_{\varphi} \delta \bar{\F} \leq \frac{v^\ast}{4}, \; \, \bar{J} + \overline{w} L_\varphi \delta \bar{\F} \leq q_1(R^\ast) \\
				&\delta^2 \bar{\Delta} \leq \frac{\bar{\nu}}{10} \delta,
			\end{split}
		\end{align}
		and $\bar{\delta}_a$ from Case 2.1, renders $x = 0$ practically semi-globally stable as per Definition~\ref{def:pract-stab} and $\hat{\theta} = \theta$ globally stable. 
	\end{IEEEproof}

	\vspace{1em}
	
	\begin{IEEEproof}(of Prop. \ref{prp:stability-subsys}) 
		The proof of Thm. \ref{thm:stability} can be transferred to a $\{I\}$-subsystem case using the new $\alpha_{\nu}$ to calculate a minimum decay rate of $\V_c$. 
		Specifically, the analysis is performed using the balls $\underline{\ball}_{\mathcal{R}}$, $\mathcal{R} \in \{R,r,R^\ast,r^\ast\}$, and new bounding functions $\alpha_{1,2}$ to relate level sets of a partition of $\V_{c}$ to the former. 
	\end{IEEEproof}
	
	\vspace{1em}

	\bibliographystyle{plain}
	\bibliography{bib/ADP_RL,bib/ClassicOptControl,bib/MPC,bib/NonlinearControl,bib/AdaptiveControl}           

\end{document}